**Spike timing regularity in vestibular afferent neurons: How ionic currents influence sensory encoding mechanisms**

Selina Baeza Loya[1]

[1]Department of Neurobiology, University of Chicago, Chicago, IL, USA

**Abstract (67 of 75 words)**
Primary vestibular neurons are categorized as either regularly or irregularly firing afferents that use rate and temporal sensory encoding strategies, respectively. While many factors influence firing in these neurons, recent work in mammalian vestibular afferents has demonstrated a rich diversity in ion channels that drive spiking regularity. Here, I review key ionic currents studied *in vitro* and demonstrate how they may enable sensory encoding strategies demonstrated *in vivo*.

**Body (2267 of 2300 words)**

**Introduction**

The peripheral vestibular organs sense head motion and tilt with respect to gravity, detecting information necessary for maintaining posture and orientation as well as providing input to construct a stable visual world. Vestibular hair cells, located in the inner ear, encode and convey sensory information to the brainstem and higher cortical areas through bipolar afferents known as vestibular ganglion neurons (VGN) (Figure 1A). The distribution of VGN firing regularity is bimodal, such that afferents are classified as either highly regular or highly irregular with respect to their spiking activity (Goldberg 2000) (Figure 1C).

Regular VGN fire at high rates, up to 400 Hz during stimulation *in vivo*. Irregular VGN fire at similar or lower rates. Such differences in firing rate can distort regularity metrics such as coefficient of variation (CV). Goldberg and colleagues therefore normalized CV by firing rate and denoted the new metric as CV* (Goldberg 2000). Regular VGN fire with high regularity with a mode CV = 0.25 or CV* = 0.05 *in vivo*. In contrast, irregular VGN fire much more erratically (CV ≥ 0.7, CV* ≥ 0.3) (Eatock et al., 2008). These differences are hypothesized to underlie two different sensory encoding strategies used by the system (Jamali et al., 2008).

First, I will discuss how ion channel composition has proved critical in driving the differences in spiking activity *in vitro*. I will then summarize *in vivo* work demonstrating regular and irregular firing correspond rate and temporal (i.e., precise spike timing) encoding strategies. Finally, I will address how the ionic mechanisms underlie differences in firing patterns in VGN in vitro can enable two parallel sensory encoding strategies to convey different aspects of head acceleration *in vivo*.

**Ionic currents in vestibular ganglion neurons modulate firing patterns**

Vestibular afferents have a bipolar morphology, with the distal branch synapsing on hair cells in the vestibular sensory epithelia and the axon projecting to the brain. VGN express a vast repertoire of ion channels that correlate with epithelial zone innervated (Eatock et al., 2008). Distinct zones (peripheral vs. central) strongly correspond to differences in spiking regularity. Regular afferents form large dendritic arbors in peripheral zones; these large arbors allow for the integration and summation of many hair cell inputs before reaching the spike initiation zone on the distal branch. In contrast, irregular afferents have a compact dendritic arbor in the central zone and contact relatively few hair cells (Figure 1A, top panel). Despite these differences, morphology alone does not drive spike timing differences as isolated VGN cell bodies still exhibit differences in spiking behavior (Goldberg 2000).

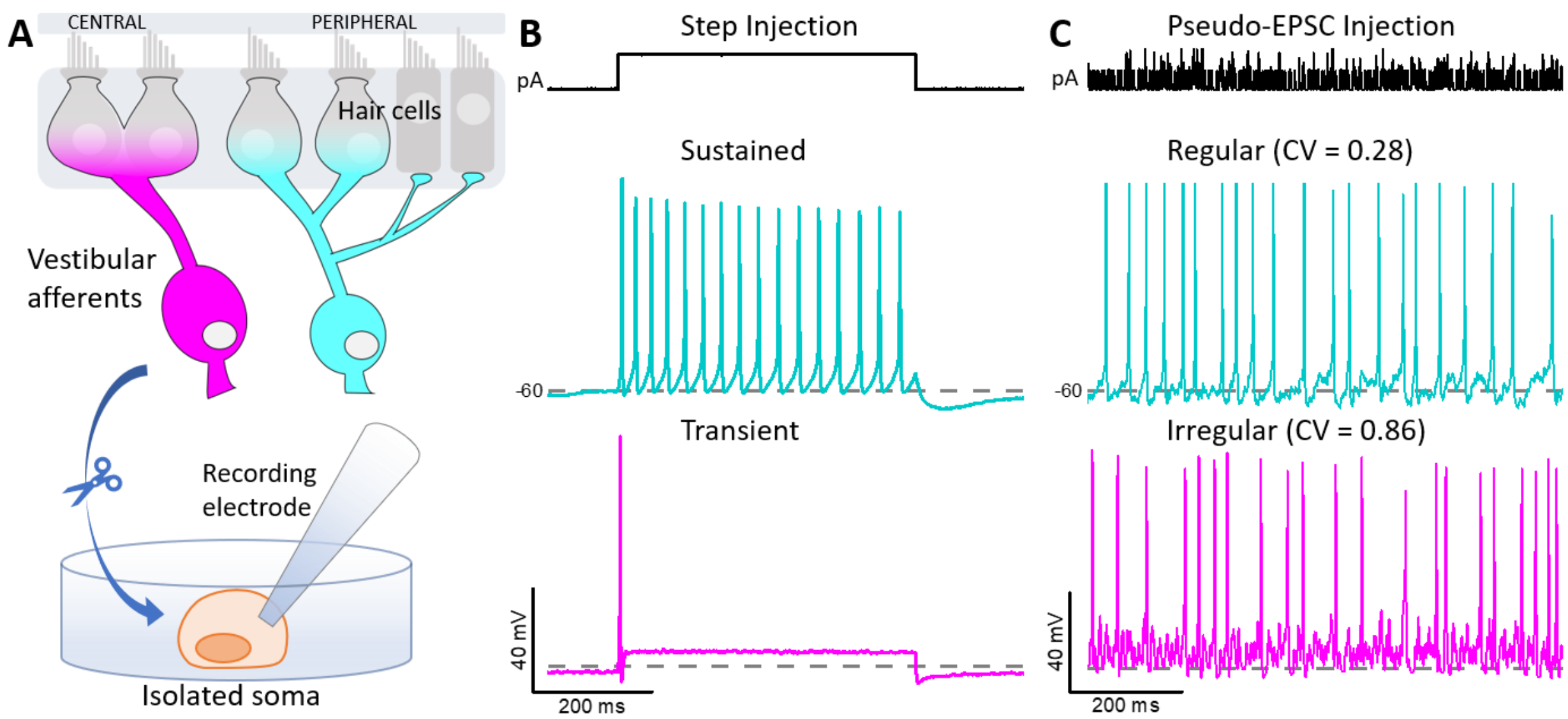

***Figure 1** Vestibular afferent neurons differ in their spiking activity.* (**A**) Top panel: Vestibular hair cells synapse onto primary vestibular afferents. The number of synaptic connections differs between zones (central vs. peripheral) of the sensory epithelia. Bottom panel: Vestibular ganglion neurons can be dissected and isolated to assess currents and spiking behaviors via electrophysiology. (**B**) Isolated VGN frequently do not fire spontaneously and must therefore be stimulated with injected current. VGN can fire with either a sustained (middle panel) or transient pattern (bottom panel) in response to a current step. (**C**) When stimulated with simulated synaptic input (pseudo-EPSCs), the same sustained VGN fires at regular intervals, whereas the transient VGN fire at irregular intervals, corresponding to vestibular afferent behavior seen in vivo.

In isolated rodent VGN somata (Figure 1A, bottom panel), regular and irregular VGN respond with sustained and transient firing patterns, respectively, to depolarizing current injections (Kalluri et al., 2010) (Figure 1B). Using this preparation, VGN were found to have voltage-gated sodium channels, calcium channels, hyperpolarization-activated cyclic nucleotide-gated (HCN) channels, calcium-gated and voltage-gated potassium channels, to name a few (Eatock et al., 2008). Sustained VGN have low current thresholds, a deep afterhyperpolarization (AHP), and fewer low voltage-activated potassium channels. Transient VGN, in contrast, have higher thresholds, a shallower AHP, and more low voltage-activated potassium channels (Eatock et al., 2008). In this review, I will focus on two types of currents, low-voltage activated potassium (I-KLV) and voltage-gated sodium (I-Na), that are hypothesized to be instrumental in spike timing regularity differences.

I-KLV has been shown to be key in producing irregular spike timing. Kalluri et al. (2010) showed, in isolated rat VGN, that the large I-KLV in transient neurons hyperpolarizes their resting membrane potential ($V_{rest}$) and decreases input resistance. This in turn increases the injected current needed to depolarize (i.e., higher current threshold). Following AP repolarization, the AHP appears truncated due to the relatively negative $V_{rest}$. A relatively hyperpolarized $V_{rest}$ allows a large spike at stimulus onset but decreases the likelihood for a continued response as I-KLV hinders subsequent spiking. Kalluri and colleagues also blocked I-KLV with α-dendrotoxin and linopirdine, indicating significant Kv1 and Kv7 contributions, and demonstrated both an increase in regularity and a decrease in spiking threshold in transient VGN. More recently, I-KLV has been shown to increase with development in both transient and sustained VGN, likely driving changes in firing patterns as VGN become more phasic after the first month of development (Ventura and Kalluri, 2019). Thus, regularity likely decreases globally as VGN mature.

Using available electrophysiological data, Hight and Kalluri (2016) developed a conductance-based VGN model to assesses the relationship between biophysical characterizations of ionic currents and spike timing. This model was later expanded by Ventura and Kalluri (2019) to include other current components such as Kv7 and hyperpolarization-activated cyclic nucleotide-gated (HCN) currents. In summary, for a transient VGN to be produced, the model must have a high level of low-voltage activated potassium conductance (gKLV, based on Kv1 and Kv7) and moderate levels of Na conductance (gNa, TTX-sensitive Na currents) as its parameters. In contrast, a fully sustained model VGN must have low gKLV and high gNa. This model not only reinforced the pivotal role of I-KLV in driving transient/irregular spiking, it also suggested that differences in I-Na may be necessary for spiking regularity.

The role of I-Na in spike timing regularity has been long overlooked due to previous reports of homogeneity in voltage dependence and tetrodotoxin (TTX) sensitivity in cultured VGN somata (e.g., Risner and Holt, 2006). However, a large I-Na may be necessary for producing the high firing rates observed in regular VGN. Additionally, I-Na may be essential for maintaining sustained firing patterns in the wake of developmentally upregulated I-KL. While work in determining the exact role of I-Na is ongoing, RT-PCR screens of rodent vestibular ganglion revealed expression of most Na channel subunits, including the TTX-insensitive $Na_V1.5$ and the TTX-resistant $Na_V1.8$. These currents were subsequently described in acutely dissociated (i.e., not cultured) rat VGN (Liu et al., 2016) and gerbil vestibular afferent endings (Meredith and Rennie, 2018). Very recent work (Rennie and Meredith, 2020) showed more I-Na current diversity in the form of persistent and resurgent Na currents, which were observed with greater frequency in mature peripheral afferent endings of regular/sustained VGN. While their impact on VGN firing is yet unknown, persistent and resurgent Na currents are known to enhance high frequency, highly regular firing in other types of sensory neurons (see Lewis and Raman, 2014 for a review).

In summary, evidence from *in vitro* studies of VGN has shown that I-KL is important for driving transient/irregular firing. Transient/irregular VGN have express large I-KL currents, and sustained/regular VGN express smaller ones. The role of these currents was reinforced via a conductance-based model, which also indicated a potential role for I-Na currents in producing sustained/regular spiking. Recent work has shown unprecedented diversity in I-Na currents, the impact of which are currently being determined.

**Spike timing regularity is strongly correlated with sensory encoding strategy**

It has long been known that regular and irregular VGN differ in their *in vivo* response dynamics (i.e., responses in time and frequency domains). Investigators have long recorded vestibular afferents in intact animals during physiological stimulation (see Goldberg 2000 for a review). To briefly summarize fifty years of work, irregular afferents are fast adapting and the gain (i.e., sensitivity) of their response increases with frequency, shown by larger responses to high-frequency head motions. Regular afferents are slower adapting and have lower sensitivity to fast head motions. Goldberg postulated that regular afferents could convey more information via their high discharge rate and low variation in firing rate. In contrast, irregular afferents have enhanced sensitivity and high intrinsic variability, both of which impact spike rate fluctuations independently of noisy input and improving information transmission (Goldberg 2000). The recent application of information theory and statistics to vestibular afferent spike trains, discussed below, has supported this idea.

Sadeghi and colleagues assessed, using information theoretic measures, how differences in spiking regularity influenced information transmission by vestibular afferents (Sadeghi et al., 2007). They applied rotational head velocities (measured in head motion frequencies) and reconstructed the stimulus from recorded vestibular afferent responses in macaques. While major conceptual findings were later revised, they showed that regular afferents were highly linear and demonstrated low variability in their responses relative to irregular afferents. Information theory metrics showed that regular afferents encoded a more significant fraction of the stimulus via linear changes in firing rate and conveyed two times more information per spike. Additionally, while irregular afferents had higher sensitivity for higher frequencies (15–20 Hz), regular afferents displayed greater relative sensitivity for lower frequencies (0.5–5 Hz). They posited that intrinsic variability (or lack thereof) in VGN spiking activity strongly influenced the encoding strategies used by vestibular afferents.

More recent work, published by Jamali and colleagues in 2016, ultimately showed how regular afferents use rate encoding and irregular afferents use precise spike timing (Jamali et al., 2016). Again, when examining changes in firing rate (i.e., the use of rate encoding), regular afferents had higher mutual information metrics relative to irregular afferents. However, when examined within the context of temporal encoding (i.e., precise spike timing), irregular afferents outperformed regular afferents. Irregular afferents displayed greater variance in their firing rate and were highly nonlinear in their responses. They showed distinctive spiking patterns to repeated trials of the same stimulus, with distinct and reproducible patterns of spikes produced within a 6-millisecond time frame. Regular afferents did not display such precise spike timing even at 30 milliseconds. Then, using an integrate-and-fire computational model of vestibular afferents to examine encoding, they varied sensitivity (likelihood of spiking) and variability (stochasticity) parameters in a model neuron. Model neurons with high levels of sensitivity and low variability increased information transmission via firing rate, like regular afferents. Neurons with high variability showed more precise spike timing performance, like irregular afferents. These *in vivo* and computational data strongly suggest that systematic differences in intrinsic variability can serve distinct forms of sensory information.

**Intrinsic variability in VGN is putatively dependent upon ionic currents**

To bridge the gap between *in vivo* and *in vitro* experiments, I posit that the intrinsic variability that influences sensory encoding stems from the balance of I-KLV and I-Na that modulates spike timing irregularity. For example, Kalluri et al. (2010) imply that transient VGN have high intrinsic variability. The transient neuron could fire a spike in response to a single, large pseudo-EPSC when spaced regularly at long equal intervals (<300 ms). However, as the intervals were systematically shortened (down to 10 ms; note that these were room temperature observations), transient VGN became less likely to fire an AP for each EPSC, assumably due to the inhibitory influence of I-KLV and the accumulation of Na channel inactivation. Therefore, even if the pseudo-EPSC input itself is regular, transient VGN still fire irregularly. This demonstrates that the variability present in isolated transient VGN does not stem from variability in injected stimuli. This constitutive irregularity may be key for producing the high intrinsic variability needed precise spike timing seen in irregular VGN *in vivo*.

To frame it in terms used by Jamali et al. 2016, while transient VGN have high intrinsic variability, sustained VGN have high sensitivity (likeliness to fire). In the same experiment discussed above in Kalluri et al. 2010, they showed sustained VGN increased in spike rate as pseudo-EPSP intervals decreased (i.e., a greater ratio of APs per number of pseudo-EPSPs) and could keep up with short intervals of 10 milliseconds. In other experiments, blocking I-KLV made transient VGN fire with greater regularity and reduced spiking threshold; not only did reducing I-KLV reduce the size of EPSCs needed to elicit firing, this block also increased the number of APs per pseudo-EPSP at smaller intervals. Thus, high I-KLV in mature transient VGN seems to correspond with increased variability, while low or moderate I-KLV in mature sustained VGN corresponds with increased sensitivity.

The precise impact of I-Na on spike timing regularity is still unknown. However, given the results available from recent studies examining these currents *in vitro* (Liu et al., 2016; Meredith and Rennie 2018, 2020) and the known kinetics of different Na currents (Raman and Lewis, 2014; Liu et al., 2016), I hypothesize that having larger I-Na would increase Na channel availability to facilitate high-frequency firing seen in regular afferents. Additionally, having I-Na currents active below or at spike threshold (e.g., persistent Na current) would increase sensitivity and probability of firing. For example, the expression of non-inactivating persistent and resurgent currents close to the spike initiation zone may enhance spiking excitability (Meredith and Rennie, 2020).

**Implications and future directions**

Vestibular afferents, and their presynaptic hair cells, are an arguably underrated yet beautiful example of sensory transmission where the encoding of a vast range of sensory information can be attributed to variation in their physiological characteristics. Information is represented in two distinct yet parallel channels that use different sensory encoding strategies to convey complementing signals. Rate encoding, used by regular vestibular afferents, transmits low-frequency head motions (<0.01 – 5 Hz), corresponding to activities such as walking and is useful in stabilizing the vestibulo-ocular reflex (Carriot et al., 2014). As seen in irregular afferents, precise spike time encoding represents head motions in the 5 – 20 Hz range, such as occur during jumping or falling (Carriot et al., 2014). Even more striking, those sensory encoding strategies seem to depend on constitutive bioelectric properties that are at the heart of differences in spike timing regularity. In addition, questions about where these afferents synapse in the vestibular brainstem, the role of efferent feedback in modulating spike timing, and how this information contributes to computations in the brain stem and cortex will also inform our understanding of vestibular sensory processing. These electrophysiological data and computational analyses all serve to inform biomedical advancements like the vestibular prosthesis and treatments of vestibular disorders.

**Citations:**

1. **Goldberg JM**. Afferent diversity and the organization of central vestibular pathways. *Exp Brain Res* 130: 277–297, 2000.
2. **Jamali M**, **Chacron MJ**, **Cullen KE**. Self-motion evokes precise spike timing in the primate vestibular system. *Nature Communications* 7: 13229, 2016.
3. **Eatock RA**, **Xue J**, **Kalluri R**. Ion channels in mammalian vestibular afferents may set regularity of firing. *Journal of Experimental Biology* 211: 1764–1774, 2008.
4. **Kalluri R**, **Xue J**, **Eatock RA**. Ion Channels Set Spike Timing Regularity of Mammalian Vestibular Afferent Neurons. *Journal of Neurophysiology* 104: 2034–2051, 2010.
5. **Ventura CM**, **Kalluri R**. Enhanced Activation of HCN Channels Reduces Excitability and Spike-Timing Regularity in Maturing Vestibular Afferent Neurons. *J Neurosci* 39: 2860–2876, 2019.
6. **Hight AE**, **Kalluri R**. A biophysical model examining the role of low-voltage-activated potassium currents in shaping the responses of vestibular ganglion neurons. *Journal of Neurophysiology* 116: 503–521, 2016.
7. **Risner JR, Holt JR.** Heterogeneous Potassium Conductances Contribute to the Diverse Firing Properties of Postnatal Mouse Vestibular Ganglion Neurons. J Neurophysiol 96: 2364–2376, 2006.
8. **Liu X-P**, **Wooltorton JRA**, **Gaboyard-Niay S**, **Yang F-C**, **Lysakowski A**, **Eatock RA**. Sodium channel diversity in the vestibular ganglion: NaV1.5, NaV1.8, and tetrodotoxin-sensitive currents. *Journal of Neurophysiology* 115: 2536–2555, 2016.
9. **Meredith FL, Rennie KJ.** Regional and Developmental Differences in Na+ Currents in Vestibular Primary Afferent Neurons. Front Cell Neurosci 12, 2018.
10. **Meredith FL**, **Rennie KJ**. Persistent and resurgent Na+ currents in vestibular calyx afferents. *Journal of Neurophysiology* 124: 510–524, 2020.
11. **Lewis AH**, **Raman IM**. Resurgent current of voltage-gated Na+ channels. *The Journal of Physiology* 592: 4825–4838, 2014.
12. **Sadeghi SG**, **Chacron MJ**, **Taylor MC**, **Cullen KE**. Neural Variability, Detection Thresholds, and Information Transmission in the Vestibular System. *J Neurosci* 27: 771–781, 2007.
13. **Carriot J**, **Jamali M**, **Chacron MJ**, **Cullen KE**. Statistics of the Vestibular Input Experienced during Natural Self-Motion: Implications for Neural Processing. *J Neurosci* 34: 8347–8357, 2014